\documentclass{template/aifrontiers}
\usepackage{template/aifrontiers}

\usepackage{booktabs}
\usepackage{array}
\usepackage{colortbl}
\usepackage{caption}
\usepackage{subcaption}
\usepackage{multirow}
\usepackage{multicol}
\usepackage{float}
\usepackage{enumitem}
\usepackage{xspace}
\usepackage{soul}
\usepackage{listings}
\usepackage[export]{adjustbox}
\usepackage{algorithm}
\usepackage{algorithmic}
\usepackage{wrapfig}
\usepackage{fontawesome5}
\usepackage{upquote}
\usepackage{textcomp}

\usepackage[T1]{fontenc}
\usepackage[utf8]{inputenc}

\usepackage[nameinlink,noabbrev]{cleveref}

\crefname{section}{Section}{Sections}
\Crefname{section}{Section}{Sections}
\crefname{subsection}{Section}{Sections}
\Crefname{subsection}{Section}{Sections}
\crefname{subsubsection}{Section}{Sections}
\Crefname{subsubsection}{Section}{Sections}

\crefname{appendix}{Appendix}{Appendices}
\Crefname{appendix}{Appendix}{Appendices}

\renewcommand{\autoref}[1]{\Cref{#1}}

\newcommand{\authmark}[1]{\textsuperscript{#1}}
\definecolor{authorcitationred}{RGB}{190,0,0}

\definecolor{peframe}{HTML}{12343B}
\definecolor{peback}{HTML}{F6F8F8}

\tcbset{pebox/.style={%
  enhanced, breakable,
  colback=peback, colframe=peframe, boxrule=0.6pt, arc=2pt,
  left=2.5mm, right=2.5mm, top=1.4mm, bottom=1.4mm,
  fonttitle=\bfseries\footnotesize, coltitle=white,
  attach boxed title to top left={xshift=2mm, yshift=-2.6mm},
  boxed title style={colback=peframe, arc=1pt, boxrule=0pt}}}

\newtcolorbox[auto counter]{promptbox}[2][]{%
  pebox, title={Box~\thetcbcounter\quad #2}, #1}

\lstdefinestyle{pe}{%
  basicstyle=\small\ttfamily, breaklines=true, columns=fullflexible,
  keepspaces=true, showstringspaces=false, upquote=true,
  aboveskip=2pt, belowskip=2pt, xleftmargin=1pt}
\newcommand{\method}{PersonaEval\xspace}

\title{\method: Persona-Based User Simulation for Evaluating Interactive Applications}
\shorttitle{\method: Persona-Based User Simulation}

\author{
\textbf{Yifan Simon Liu}\thanks{Equal contribution.},
\textbf{Qianfeng Wen}\footnotemark[1],
\textbf{Yilan Fan}\footnotemark[1],
\textbf{Shirley Huang}\footnotemark[1],
\textbf{Ruoqi Gao}\footnotemark[1],
\textbf{Jianheng Hou}\footnotemark[1],
\\[0.55em]
Muhammad Ahmed Mohsin\thanks{Core execution team.},
Zonglin Di\footnotemark[2],
Brihi Joshi\footnotemark[2],
Xincheng Tan\footnotemark[2],
Yucheng Lu\footnotemark[2],
\\
Xiaoyi Liu\footnotemark[2],
Heming Liu\footnotemark[2],
Hanwen Xing\footnotemark[2],
Guanghui Min\footnotemark[2],
Zhengyang Shan\footnotemark[2],
\\
My Chiffon Nguyen\footnotemark[2],
Ishan Gupta\footnotemark[2],
\\[0.55em]
Yunze Xiao,
Hannah Collison,
Jintao Huang,
Jiatong Li,
Sankalp Jajee,
Yunhan Zhao,
\\
Bing Hu,
Sky Ng,
Xupeng Chen,
Binghang Lu,
Weihang Xiao,
Aravind Mohan,
\\
Bolun Sun,
Yunshu Wu,
Yuanda Xu,
Yun Shen,
Runyu Zhang,
Zheyuan Deng,
\\
Zhiwei Zhang,
Qianyu Zhu,
Dianzhuo Wang,
Yijun Wang,
Yixuan He,
\\[0.55em]
\textbf{Yuexing Hao}\thanks{Team leaders and corresponding authors.},
\textbf{Xiaomin Li}\footnotemark[3]
\\[0.55em]
{\normalsize MatrAIx Team}\\
{\normalsize \texttt{info@matraix.ai}}
}

\date{\today}

\renewcommand{\weblink}{}
\renewcommand{\foundrylink}{}
\renewcommand{\hflink}{}
\renewcommand{\ghlink}{}

\renewcommand{\AIFLogo}{%
  \raisebox{-0.1ex}{%
    \includegraphics[height=1.5em]{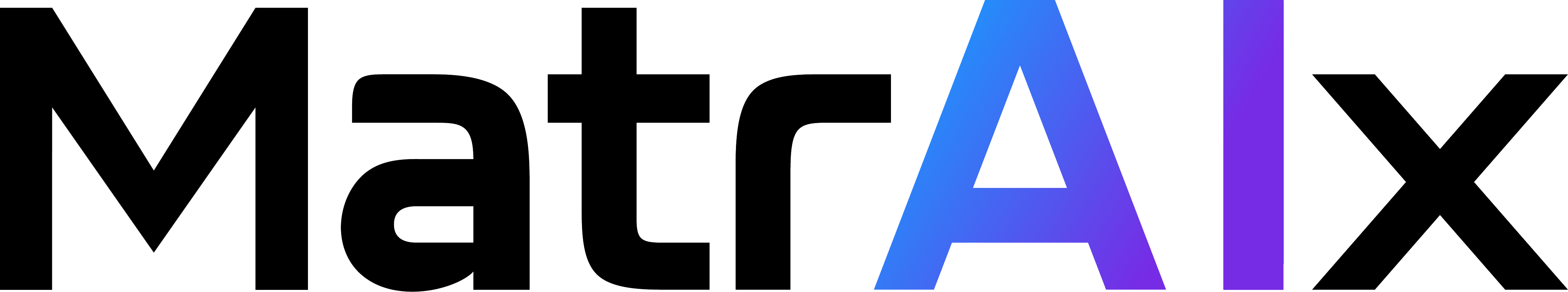}}%
}

\begin{document}

\begin{abstract} 
Real user studies are important for understanding how people interact with systems under test or already deployed. In practice, however, they are often costly, time-consuming, and difficult to scale. To address these challenges, we introduce \method, a persona-based user simulation framework that approximates real-user behavior across diverse interactive settings. \method connects simulated users drawn from existing persona datasets to task-specific application interfaces and collects the interaction trajectories and outcomes. 
\method  provides a plug-and-play evaluation workflow in which the application being evaluated can be easily changed. In this demo, we present \method on three forms of interactive applications: surveys, 
chatbots, 
and web applications.
Together, these examples show that \method can support repeatable, parallelizable, and scalable evaluation across different interaction settings, while producing user-oriented feedback and task-specific behavior.
\end{abstract}

\maketitle

\section{Introduction}

Interactive applications can produce different outcomes across users because they bring different goals and preferences to the same system.  Understanding this variation is important during early development. Real user studies remain the most reliable way to measure human experience, but their cost, recruitment burden, and slow iteration cycle make them difficult to use for rapid evaluation~\citep{xuan2025gidea,lu2025uxagent}.

Persona-based simulation offers a practical complement by simulating users with specified profiles and having them interact with a target system. Recent work has developed richer persona representations~\citep{zhang2018personachat,ge2024personahub,nvidia2025nemotron,deeppersona2025}, while LLM-based agent frameworks make it easier to deploy these personas as interactive agents~\citep{park2023generative,vezhnevets2023concordia,yang2024oasis,microsoft2024tinytroupe}. However, many existing pipelines are designed for a single task format, limiting the reuse of the same persona population across systems and interaction settings.

Therefore, we introduce \method (\autoref{fig:personaeval}) as a plug-and-play system for persona-based user simulation for evaluating interactive applications. Our contributions are as follows:
\begin{itemize}
    \item \textbf{A plug-and-play persona simulation system.} We build \method \footnote{\href{https://github.com/YifanLiu2/persona-eval.git}{Code} and \href{https://github.com/YifanLiu2/persona-eval.git}{demo video} uploaded.} as a modular system that connects persona-based simulated users to application adapters, allowing developers to plug different personas into different interactive systems while collecting interaction trajectories and task outcomes in a common format.

    \item \textbf{A multi-application demonstration.} We instantiate \method on survey, chatbot, and web settings, demonstrating that our system can support evaluation across different interaction settings.
\end{itemize}
\section{Related Work}
 \label{sec:related-personas} 

\paragraph{Persona datasets.} Persona-grounded modeling has evolved from small manually written profiles such as PersonaChat~\citep{zhang2018personachat} to large-scale synthetic persona datasets. Recent efforts have generated millions to billions of personas from web data, demographic priors, and LLM synthesis, including Persona-Hub~\citep{ge2024personahub}, NVIDIA Nemotron-Personas~\citep{nvidia2025nemotron}, and DeepPersona~\citep{deeppersona2025}. Parallel work has developed methods for persona-conditioned dialogue and behavior generation through synthetic data and instruction tuning~\citep{jandaghi2023syntheticpersonachat,wang2025opencharacter,wang2023selfinstruct}. 
\paragraph{Persona simulation and evaluation.} Persona-grounded agents have enabled increasingly realistic simulations through frameworks such as Concordia~\citep{vezhnevets2023concordia}, OASIS~\citep{yang2024oasis}, TinyTroupe~\citep{microsoft2024tinytroupe}, and AI Town~\citep{a16z2023aitown}. Parallel work evaluates whether agents faithfully follow assigned personas using benchmarks such as PersonaGym, CharacterEval, SimBench, and $\tau$-bench~\citep{samuel2024personagym,tu2024charactereval,simbench2025,yao2024taubench,li2026matraix,lu2026personagrounding,ngmicroverse}. Our work is complementary to these efforts by providing a modular simulation platform that conditions recommendation agents on arbitrary persona profiles, enabling systematic evaluation across diverse personas and applications.


\section{System Description}
\begin{figure}
    \centering
    \includegraphics[width=0.98\linewidth]{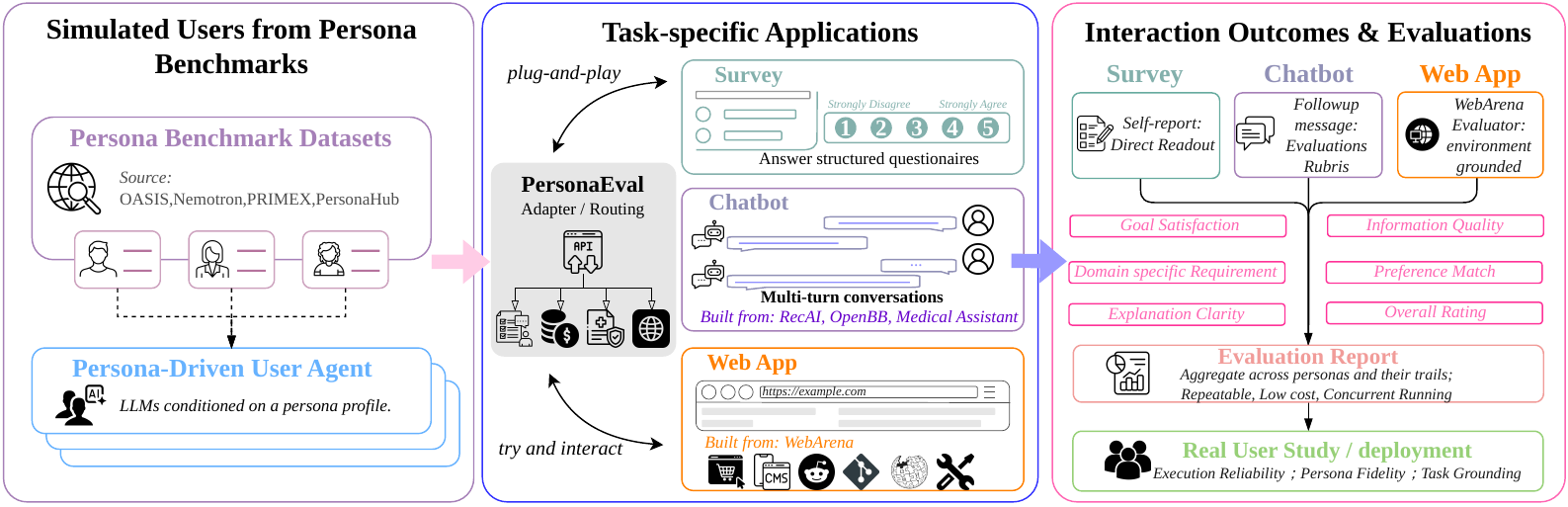}
    \caption{
    \textbf{Overview of \method.}
    \method connects persona-driven simulated users to different types of applications through plug-and-play adapters and provides a unified workflow for collecting interaction outcomes, feedback, and evaluation reports.
    }
    \label{fig:personaeval}
\end{figure}

\method(\autoref{fig:personaeval}) is a system for evaluating interactive applications through simulations of realistic users. Each simulation uses a simulated user with predefined persona attributes, such as background, preferences, and communication style. This lets \method test how the same application is experienced by different kinds of users. 

Each simulation run is executed inside a sandbox with declared resources and controlled dependencies, which provides the runtime boundary for the evaluation and keeps the persona side and the task side isolated from other applications. This is important because different applications may need different code, data, tools, browser environments, or backend services.

\paragraph{Persona API.} The persona API wraps a persona model as a simulated user. It receives the assigned persona and the task instruction, then produces the next user behavior through the interface required by the task. Depending on the application type, this behavior can be a survey response, a chat message, or a computer-use action. 

\paragraph{Task API.} The task API packages the environment needed to test a particular application. It defines the task instruction, hosts or connects to the application under test, manages the interaction protocol, and provides the evaluation form. For example, a chatbot task connects the simulated user to a chat API, and a web task hosts a website and exposes it through a browser environment. This design lets each task define its own environment and evaluation criteria while sharing the same simulation workflow.

\section{Applications}
\label{sec:applications}

\begin{figure*}[t]
    \centering
    \includegraphics[width=0.98\textwidth]{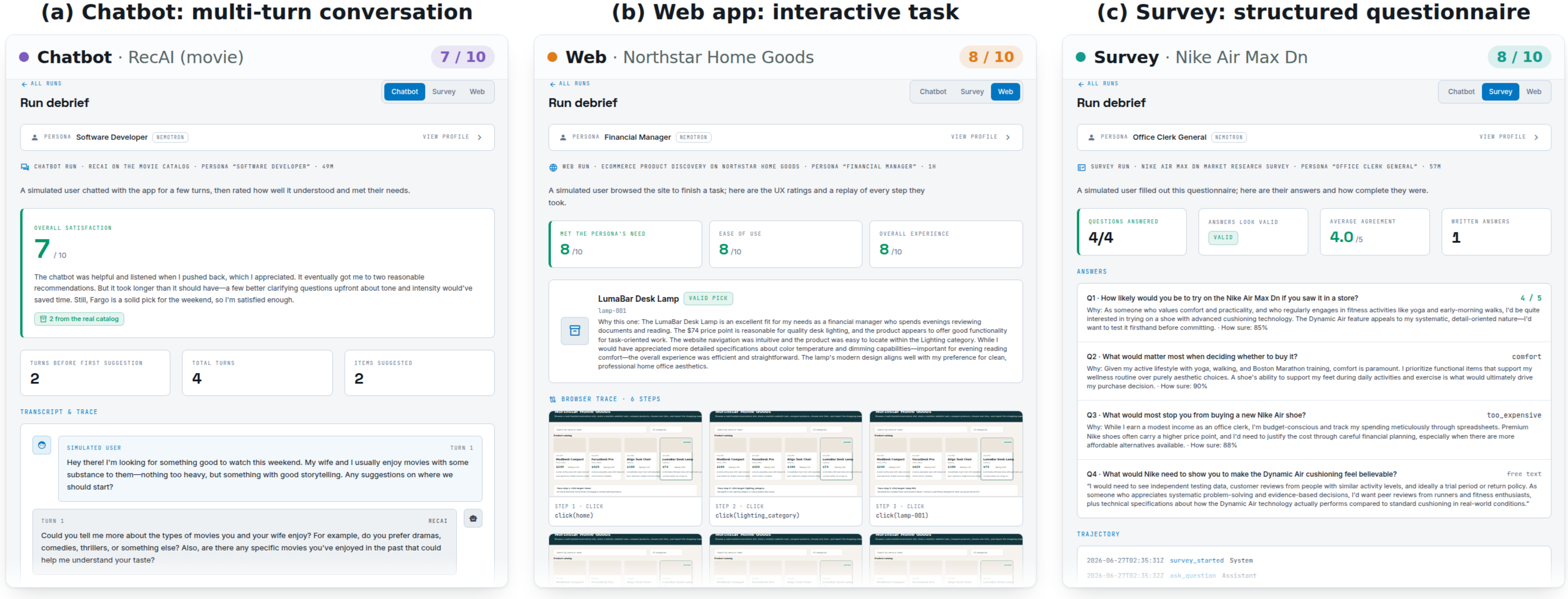}
    \caption{
    \textbf{Application demos in \method.}
    \method supports structured survey responses, multi-turn chatbot conversations, and browser-based web tasks under the same persona-driven evaluation workflow.
    }
    \label{fig:application-demo}
\end{figure*}

\method evaluates applications through task adapters. Each adapter specifies what the simulated user should do and how the interaction is evaluated. As shown in \autoref{fig:application-demo}, our demo organizes applications into three types: survey, chatbot, and web.

\paragraph{Survey.}
A survey task asks a simulated user to read a product brief or task background and complete a structured questionnaire. The evaluation is based on the completed survey responses, which record the user's opinion, preference, and explanation.

\paragraph{Chatbot.}
A chatbot task asks a simulated user to complete a realistic goal through a multi-turn conversation. We use recommendation chatbots for movies and beauty products based on RecAI and InteRecAgent~\citep{huang2023recommender,lian2024recai}, a financial research chatbot based on OpenBB~\citep{openbb,openbbmcp}, and a medical consultation chatbot based on a multi-agent medical assistant~\citep{majumder2025medicalassistant}. The evaluation is based on a post-interaction form completed by the simulated user, covering goal satisfaction, constraint following, preference match, clarification quality, overall rating, and the reason for the rating.

\paragraph{Web.}
A web task asks a simulated user to complete a browser-based goal, such as searching, comparing, and selecting a product on a WebArena ecommerce site~\citep{zhou2024webarena}. The evaluation records whether the user completed the task, what outcome they selected, how easy the website was to use, the overall rating, and the reason for that rating.

Detailed task description and evaluation forms are provided in \autoref{app:adapters}.

\section{Results}
\label{sec:results}

\begin{figure*}[t]
  \centering
  \includegraphics[width=0.90\textwidth]{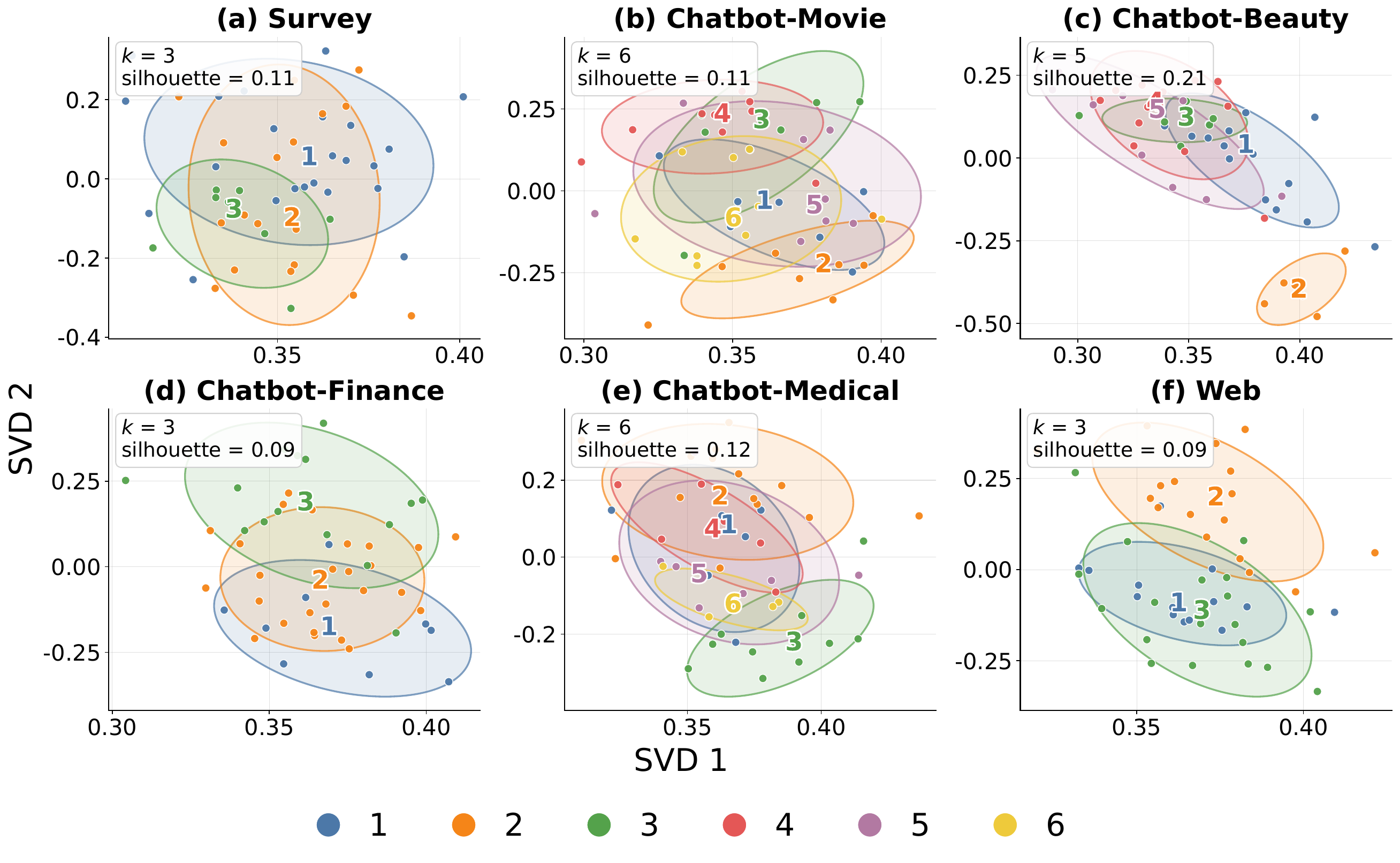}
    \caption{  \textbf{Diversity of selected personas.}
  For each application instance, the 50 selected personas are clustered in a TF--IDF$\rightarrow$SVD space.}
  \label{fig:within}
\end{figure*}

\paragraph{Setup.}
For each application, we select 50 relevant Nemotron personas~\citep{nvidia2025nemotron} by embedding similarity between the application description and persona profiles to approximate the selection of real users whose backgrounds match the target application. As shown in \autoref{fig:within}, the selected personas retain within-cluster diversity in each application domain. Details of persona selection are provided in \autoref{app:diversity}. We use \texttt{Claude Haiku 4.5}~\citep{anthropic2025haiku45} for persona-driven users, \texttt{GPT-4o mini}~\citep{openai2024gpt4omini} for chatbot applications, and cap chatbot conversations at eight turns.

\begin{figure}[t]
    \centering
    \includegraphics[width=0.80\linewidth]{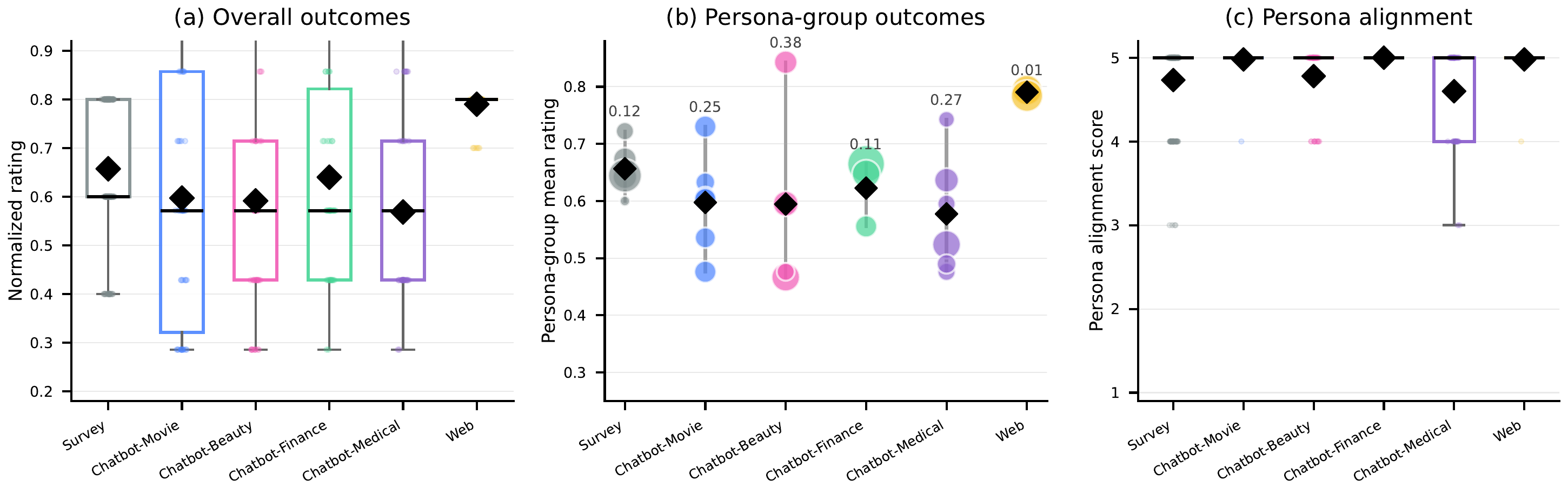}
    \caption{
        \textbf{Results overview across survey, chatbot, and web settings.}
        Panel (a) shows normalized overall ratings, with diamonds marking application means. Panel (b) shows persona-group mean ratings; point size is proportional to group size, and numbers show the max--min group spread. Panel (c) shows persona alignment score distributions.
    }
    \label{fig:results-components}
\end{figure}

\paragraph{Overall result analysis.}
\autoref{fig:results-components}(a) analyzes simulated users' reported overall ratings. The chatbot tasks show moderate satisfaction with wider spread, because personas bring different goals, preferences, and domain knowledge into the interaction. Survey and web results are more concentrated, suggesting that structured survey responses and constrained web checkout produce more stable outcomes. Detailed results and representative interaction traces are provided in \autoref{app:results}.

\begin{figure}[t]
    \centering
    \includegraphics[width=0.98\linewidth]{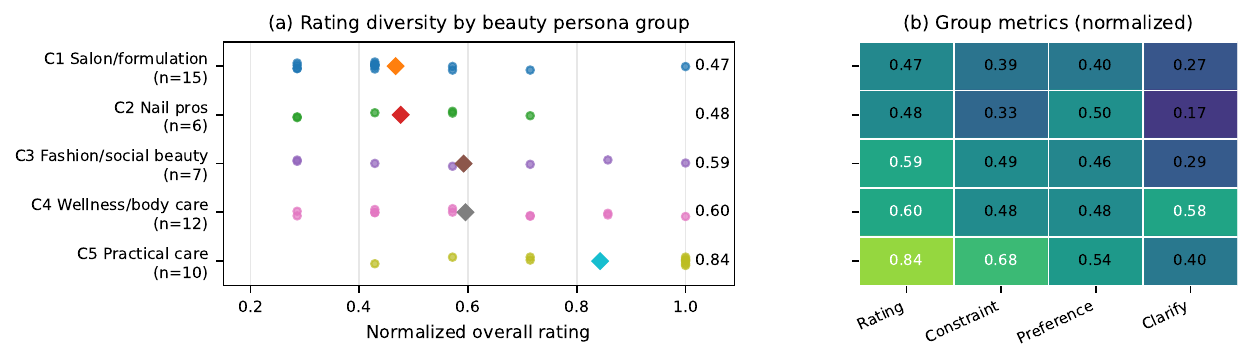}
    \caption{
    \textbf{Beauty persona group results.}
    We group beauty users by shared persona traits: C1 salon and formulation users, C2 nail professionals, C3 fashion and social beauty users, C4 wellness and body care users, and C5 practical care users. Panel (a) shows normalized ratings for each persona, with diamonds marking group means. Panel (b) summarizes normalized group metrics. 
    }
    \label{fig:beauty-groups}
\end{figure}

\paragraph{Behavior differences across personas.}
\autoref{fig:results-components}(b) summarizes outcome variation across persona groups. Larger spreads in beauty, medical, and movie suggest that different user groups evaluate the same application differently, while survey, finance, and web show more similar group ratings. \autoref{fig:beauty-groups} shows the beauty case in more detail. We read the full persona profiles to name each group. Salon and formulation users (C1) and nail professionals (C2) give lower ratings because their goals often require professional sourcing, ingredient details, or salon inventory support. Practical care users (C5) give higher ratings because their needs better match the available product catalog.

\paragraph{Persona alignment analysis.}
We evaluate persona alignment with a human-judged 1--5 score that measures whether each simulated interaction is consistent with the assigned user persona. The judge considers the user's task goal, interaction trace, and final evaluation form. \autoref{fig:results-components}(c) shows that persona alignment remains high across survey, chatbot, and web applications. Lower scores appear more often in tasks where users need to express professional knowledge, personal constraints, or domain-specific concerns over multiple turns. Overall, these results suggest that \method produces persona-aligned simulations.

\section{Conclusion}
\label{sec:conclusion}

We presented \method, a plug-and-play system for persona-based user simulation for evaluating interactive applications. Across survey, chatbot, and web settings, \method shows application-level differences, persona-group variation, and persona alignment patterns. These findings position \method as a useful tool for simulating real-user studies of interactive systems. Future work should calibrate simulations against human data and expand to broader domain of applications.

\paragraph{Limitations and Future Work.}
Our current demo covers a limited set of survey, chatbot, and web applications. Future work will expand the application coverage and conduct more fine-grained analyses of persona-driven behavior. We also plan to validate simulation quality more rigorously through comparisons with real-user behavior, human evaluation of persona alignment, and deeper analysis of whether users' goals, interactions, and evaluations are consistently grounded in their assigned personas.

\clearpage
\bibliographystyle{plainnat}
\bibliography{references}

\appendix
\section{Authors}
\label[appendix]{app:authors}

Author names and their affiliation details are provided here. Affiliation
superscripts refer to the numbered list of institutions at the end of this
section.

\subsection{Contributor roles and affiliations}

\noindent\textbf{Core Execution Team.}
Yifan Simon Liu\authmark{*,11},
Qianfeng Wen\authmark{*,11},
Yilan Fan\authmark{*,13},
Shirley Huang\authmark{*,5},
Ruoqi Gao\authmark{*,12},
Jianheng Hou\authmark{*,4},
Muhammad Ahmed Mohsin\authmark{12},
Zonglin Di\authmark{10},
Brihi Joshi\authmark{4},
Xincheng Tan\authmark{8},
Yucheng Lu\authmark{1},
Xiaoyi Liu\authmark{2},
Heming Liu\authmark{3},
Hanwen Xing\authmark{4},
Guanghui Min\authmark{6},
Zhengyang Shan\authmark{7},
My Chiffon Nguyen\authmark{8},
and Ishan Gupta\authmark{9}.

\smallskip
\noindent\textsuperscript{*} denotes equal contribution.

\medskip
\noindent\textbf{Contributors.}
Yunze Xiao\authmark{28},
Hannah Collison\authmark{15},
Jintao Huang\authmark{16},
Jiatong Li\authmark{17},
Sankalp Jajee\authmark{18},
Yunhan Zhao\authmark{19},
Bing Hu\authmark{26},
Sky Ng\authmark{8},
Xupeng Chen\authmark{1},
Binghang Lu\authmark{21},
Weihang Xiao\authmark{22},
Aravind Mohan\authmark{23},
Bolun Sun\authmark{14},
Yunshu Wu\authmark{8},
Yuanda Xu\authmark{24},
Yun Shen\authmark{16},
Runyu Zhang\authmark{25},
Zheyuan Deng\authmark{2},
Zhiwei Zhang\authmark{20},
Qianyu Zhu\authmark{25},
Dianzhuo Wang\authmark{5},
Yijun Wang\authmark{5},
and Yixuan He\authmark{27}.

\medskip
\noindent\textbf{Team Leadership \& Correspondence.}
Yuexing Hao\authmark{\(\dagger\),25} and
Xiaomin Li\authmark{\(\dagger\),5}. \\
\textsuperscript{\(\dagger\)} Team leaders and corresponding authors.
Correspondence:
\href{mailto:yuexing@mit.edu}{\texttt{yuexing@mit.edu}}
and
\href{mailto:xiaominli@g.harvard.edu}{\texttt{xiaominli@g.harvard.edu}}.

\subsection*{Affiliations}
\begingroup
\small
\setlength{\columnsep}{1.5em}
\begin{multicols}{2}
\begin{enumerate}[
    leftmargin=*,
    labelsep=0.45em,
    itemsep=0.28em,
    parsep=0pt,
    topsep=0.25em
]
  \item New York University
  \item Brown University
  \item University of Illinois Urbana-Champaign
  \item University of Southern California
  \item Harvard University
  \item University of Virginia
  \item Boston University
  \item Independent Contributor
  \item University of California, San Diego
  \item University of California, Santa Cruz
  \item University of Toronto
  \item Stanford University
  \item Georgia Institute of Technology
  \item Northwestern University
  \item Johns Hopkins University
  \item The Ohio State University
  \item University of Wisconsin--Madison
  \item Medical University of South Carolina
  \item University of California, Irvine
  \item Pennsylvania State University
  \item Purdue University
  \item Cornell University
  \item University at Buffalo
  \item Princeton University
  \item Massachusetts Institute of Technology
  \item University of California, Riverside
  \item Arizona State University
  \item Carnegie Mellon University
\end{enumerate}
\end{multicols}
\endgroup
\section{Extended Related Work}

\paragraph{Large language models and agents.} Large language models (LLMs) exhibit capabilities in multi-step reasoning \citep{wei2022chain,wen2024mcts,wen2025chessqa,tang2026groundedchess,jiao2026thinktwice} and role-conditioned generation, allowing them to produce behavior consistent with specified characters or personas~\citep{shanahan2023role}. When augmented with memory, planning, and environment interaction, LLMs can further serve as agents that reason, act, and adapt over multi-step trajectories~\citep{yao2023react,park2023generative}.

\paragraph{Persona datasets.} Persona-grounded modeling has evolved from small manually written profiles such as PersonaChat~\citep{zhang2018personachat} to large-scale synthetic persona datasets. Recent efforts have generated millions to billions of personas from web data, demographic priors, and LLM synthesis, including Persona-Hub~\citep{ge2024personahub}, NVIDIA Nemotron-Personas~\citep{nvidia2025nemotron}, and DeepPersona~\citep{deeppersona2025}. Parallel work has developed methods for persona-conditioned dialogue and behavior generation through synthetic data and instruction tuning~\citep{jandaghi2023syntheticpersonachat,wang2025opencharacter,wang2023selfinstruct}. 

\paragraph{Persona simulation and evaluation.} Persona-grounded agents have enabled increasingly realistic simulations through frameworks such as Concordia~\citep{vezhnevets2023concordia}, OASIS~\citep{yang2024oasis}, TinyTroupe~\citep{microsoft2024tinytroupe}, and AI Town~\citep{a16z2023aitown}. Parallel work evaluates whether agents faithfully follow assigned personas using benchmarks such as PersonaGym, CharacterEval, SimBench, and $\tau$-bench~\citep{samuel2024personagym,tu2024charactereval,simbench2025,yao2024taubench}. Our work is complementary to these efforts by providing a modular simulation platform that conditions recommendation agents on arbitrary persona profiles, enabling systematic evaluation across diverse personas and applications.

\paragraph{Applications: survey, chatbot, and web.} Persona agents are increasingly deployed as proxies for human respondents. In the \emph{survey} setting, an agent reads a stimulus and returns structured feedback: large-scale replications find that LLMs recover 73--81\% of main effects from human experiments, albeit with inflated effect sizes and weaker fidelity on sensitive topics~\citep{Cui+2024}. In the \emph{chatbot} setting, an agent converses with a target system and evaluates it, most developed for conversational recommendation~\citep{He+2023,wen2024elaborative,wen2025elaborative,liu2026multimodal,wen2026safegeo} and RAG-based conversational agents \citep{liu2025madpr,kim2026bagel,liu2026semanticxpath,liang2026goalmem,liu2026segtreemem,liu2025medievallatin}, and the central shift is toward interactive, simulator-driven evaluation, since static single-turn protocols substantially underestimate quality~\citep{Wang+2023,liang2026insitu} and simulators exhibit failure modes such as popularity bias~\citep{Yoon+2024}. In the \emph{web} setting, agents probe products and interfaces before any human study, generating thousands of testers for live websites~\citep{lu2025uxagent}, populating prototype social spaces~\citep{Park+2022}, and running A/B tests on real shopping interfaces~\citep{lu2026agentab}.

\section{\method Construction Details}
\label[appendix]{app:construction}

This appendix documents how the \method demo is built: the persona corpus and
how applications are matched to personas (\autoref{app:personas}), the simulated
user exposed by the Persona API (\autoref{app:simulator}), the three task adapters
and their interaction protocols (\autoref{app:adapters}), and the models and run
settings behind the results in 5 (\autoref{app:settings}). 

\subsection{Persona Corpus and Selection}
\label[appendix]{app:personas}

\method treats a persona as a fixed, self-contained description of one user.
Personas are drawn from existing datasets rather than authored per task, so the
same population can be reused across applications. The demo bundles 336 curated
profiles from four sources (Nemotron-Personas-USA, PersonaHub, PRIMEX, and
OASIS \citep{nvidia2025nemotron, ge2024personahub, appleprimex, yang2024oasis}); the runs reported in \autoref{sec:results} use the Nemotron-Personas-USA subset. Each
profile is a structured record --- demographics, a set of facet narratives (for
example professional, culinary, and social facets), a free-text background, and
explicit attribute lists --- that the loader renders into a single humanized
text block (Box~\ref{box:persona}). 

\paragraph{Selection.} For each application we embed its short description and
rank personas by similarity between that description and each profile, keeping
the 50 nearest. This approximates recruiting users whose background matches the
application, and gives every application a 50-persona test set without manual
curation.

\begin{promptbox}[label={box:persona}]{Persona profile record (rendered to one text block)}
\begin{lstlisting}
id: <hash>                 source: Nemotron      # Nemotron | PersonaHub | PRIMEX | OASIS
demographics:
  age, gender, marital_status, education_level, occupation
  location: { city, state, country, zipcode }
personas:                  # facet narratives, a few sentences each
  professional, sports, arts, travel, culinary, core
background:                # free-text life context (e.g. cultural)
attributes:
  skills: [...], hobbies: [...], career_goals: ...

# The loader humanizes the keys and renders the record into ONE text block,
# used both in the persona drawer and verbatim as the {persona_profile} below.
\end{lstlisting}
\end{promptbox}

\subsection{Simulated User (Persona API)}
\label[appendix]{app:simulator}

The Persona API wraps a single language model as one simulated user. Its system
prompt composes the application context, a short description of the application
under test, and the rendered persona block; a goal-context template then directs
the user to (i) infer a realistic goal and the constraints or preferences that
matter to it, (ii) reveal those needs gradually rather than all at once,
(iii) react to the application by confirming, pushing back, or asking for
clarification, and (iv) keep messages short (one to three sentences).
Box~\ref{box:sysprompt} gives the template. The model emits strict JSON at every
step (JSON-object decoding, temperature 0.7), and the same persona-conditioned
instance later completes the post-interaction form in character, so the
evaluation reflects the assigned user's point of view rather than an external
judge.

\begin{promptbox}[label={box:sysprompt}]{Simulated-user system prompt}
\begin{lstlisting}
You are a real user of this interactive application.

Application context: {domain}

{application_description}

Your assigned persona (stay in character at all times):
{persona_profile}

Based on your assigned persona, first decide what realistic goal you want to
accomplish with this application and what constraints or preferences matter
most to you. Then behave like a genuine human user:
- Do NOT reveal everything at once: share your needs gradually, as a real
  person would, and answer the agent's follow-up questions naturally.
- React to the application's responses: if they fit your needs, say so; if
  not, push back, refine, or ask for clarification.
- Keep messages short and conversational (1-3 sentences).
\end{lstlisting}
\end{promptbox}

\subsection{Interaction Protocols and Task Adapters}
\label[appendix]{app:adapters}

Each application is wrapped as a task that fixes an interaction protocol, an
interface, and an evaluation form. The persona side is shared by all tasks, while
the adapter differs across survey, chatbot, and web applications. All JSON
schema field names below use camelCase; fixed identifiers such as survey question
ids, application keys, product ids, and option ids are treated as data values and
are kept as written.

\paragraph{Survey.}
The protocol is a single survey submission. The simulated user receives a market
research instrument and returns one JSON object with answers, rationales,
confidence scores, and a trajectory. The evaluation is the completed survey
itself plus validation checks.

\begin{promptbox}[label={box:survey-task}]{Survey adapter: task prompt}
\begin{lstlisting}
You are completing a market research survey.

Read the survey context and answer each question as the assigned persona.
Return one JSON object.

Survey context:
{
  "instrumentId": "<instrument id>",
  "title": "<survey title>",
  "concept": "<survey concept>",
  "questions": [ ... ]
}
\end{lstlisting}
\end{promptbox}

\begin{promptbox}[label={box:survey-schema}]{Survey adapter: response schema and validation}
\begin{lstlisting}
# Response schema
{
  "instrument": {
    "instrumentId": "<instrument id>",
    "title": "<survey title>"
  },
  "answers": [
    {
      "questionId": "<question id>",
      "answerValue": "<likert number, choice string/list, or free text>",
      "answerRationale": "<short persona-grounded reason>",
      "confidence": 0.0
    }
  ],
  "trajectory": [
    {
      "timestamp": "2026-06-24T00:00:00Z",
      "actor": "user",
      "eventType": "answerQuestion",
      "context": {
        "questionId": "<question id>"
      },
      "outcome": {
        "questionId": "<question id>",
        "answerValue": 4
      }
    }
  ]
}

# Validation
- all required questions are answered
- question ids are valid
- Likert values are within the valid range
- choice values match the allowed options
- free-text answers are non-empty
- trajectory contains started, ask, answer, and completed events
- summary metrics include numQuestions, numAnswered, and meanLikert
\end{lstlisting}
\end{promptbox}

\begin{promptbox}[label={box:survey-instruments}]{Survey adapter: demo instruments}
\begin{lstlisting}
# Instrument 1
instrumentId: chatgpt_images_market_research_v1
title: ChatGPT Images Market Research Survey
concept: ChatGPT image generation and editing
questions:
  - questionId: trial_intent
    questionType: likert
    scale: 1-5
    prompt: likelihood of using ChatGPT to create or edit images

  - questionId: most_useful_task
    questionType: singleChoice
    options:
      - creating_social_media_images
      - editing_or_fixing_photos
      - making_presentations_or_posters
      - brainstorming_visual_ideas
      - none_of_these

  - questionId: adoption_barrier
    questionType: singleChoice
    options:
      - image_does_not_match_my_request
      - concerns_about_photo_privacy
      - hard_to_get_realistic_results
      - prefer_existing_design_tools
      - not_sure_when_to_use_it

  - questionId: regular_use_trigger
    questionType: freeText


# Instrument 2
instrumentId: instagram_reels_market_research_v1
title: Instagram Reels Market Research Survey
concept: short-video watching, creating, and discovery
questions:
  - questionId: watch_intent
    questionType: likert
    scale: 1-5
    prompt: likelihood of watching Reels during a normal Instagram visit

  - questionId: content_pull
    questionType: singleChoice
    options:
      - funny_or_entertaining_videos
      - friends_or_creators_i_follow
      - how_to_or_learning_content
      - news_trends_or_pop_culture
      - shopping_or_product_discovery
      - none_of_these

  - questionId: usage_barrier
    questionType: singleChoice
    options:
      - too_many_ads
      - irrelevant_recommendations
      - too_addictive_or_time_wasting
      - low_quality_content
      - privacy_or_tracking_concerns

  - questionId: improvement_request
    questionType: freeText


# Instrument 3
instrumentId: nike_air_max_dn_market_research_v1
title: Nike Air Max Dn Market Research Survey
concept: Nike Air Max Dn and Dynamic Air cushioning
questions:
  - questionId: try_on_intent
    questionType: likert
    scale: 1-5
    prompt: likelihood of trying it on in store

  - questionId: purchase_driver
    questionType: singleChoice
    options:
      - comfort
      - style
      - price
      - brand_reputation
      - durability

  - questionId: adoption_barrier
    questionType: singleChoice
    options:
      - too_expensive
      - not_my_style
      - unsure_about_comfort
      - already_have_similar_shoes
      - prefer_other_brands

  - questionId: proof_needed
    questionType: freeText
\end{lstlisting}
\end{promptbox}

\paragraph{Chatbot.}
The protocol is a multi-turn conversation. The simulated user receives a chatbot
description, forms a realistic goal based on the assigned persona, reveals needs
gradually, answers follow-up questions, and stops when they can judge whether the
chatbot satisfied the need. The recommendation tasks
use RecAI and InteRecAgent~\citep{huang2023recommender,lian2024recai}; the
finance task uses OpenBB~\citep{openbb,openbbmcp}; and the medical task uses a
multi-agent medical assistant~\citep{majumder2025medicalassistant}.

\begin{promptbox}[label={box:chat-task}]{Chatbot adapter: task prompt}
\begin{lstlisting}
You are a user of a <systemLabel>.

System description:
<systemDescription>

Context for this interaction:
Based on your assigned persona, silently decide what you realistically
want from this system and which constraints or preferences matter to you.
Start the conversation naturally. Do not reveal everything at once.
Let the system ask follow-up questions, answer in character, and give
feedback when a response does not fit. Continue until you can judge
whether the system satisfied your need.
\end{lstlisting}
\end{promptbox}

\begin{promptbox}[label={box:chat-schema}]{Chatbot adapter: next-turn schema and key rules}
\begin{lstlisting}
# Next-turn response schema
{
  "message": "string|null",
  "done": "boolean",
  "doneReason": "string|null"
}

# Key rules
- stay in character
- decide a realistic need from the assigned persona
- reveal preferences gradually
- set done=true only when the user need is already satisfied
- if continuing, keep the message short and natural
\end{lstlisting}
\end{promptbox}

\begin{promptbox}[label={box:chat-eval}]{Chatbot adapter: evaluation form}
\begin{lstlisting}
# Post-interaction evaluation form
{
  "constraintSatisfaction":
    "<1-5 score for how well the user's need and constraints were met>",
  "constraintRationale":
    "<short reason>",

  "preferenceSatisfaction":
    "<1-5 score for how well personal preferences were met>",
  "preferenceRationale":
    "<short reason>",

  "overallRating":
    "<integer score on the task-specific rating scale>",
  "ratingReason":
    "<short reason for the rating>",

  "clarificationUseful":
    "<true|false>",
  "clarificationRationale":
    "<which questions were useful, or why not>"
}
\end{lstlisting}
\end{promptbox}

The finance and medical tasks currently use the same shared chatbot evaluation
form rather than separate domain-specific scoring fields. The system descriptions
used by the chatbot adapter are listed below.

\begin{promptbox}[label={box:chat-apps}]{Chatbot adapter: applications under test}
\begin{lstlisting}
# Movie recommendation
applicationKey: recai:movie
systemLabel: movie recommendation chatbot
systemDescription:
  A conversational movie recommender that asks about tastes and constraints,
  searches a film catalog, and suggests movies.

# Beauty recommendation
applicationKey: recai:beauty_product
systemLabel: beauty product recommendation chatbot
systemDescription:
  A beauty and personal-care recommender that asks about needs, skin type,
  budget, and preferences, then searches a beauty catalog.

# Financial research
applicationKey: finance_openbb:financial_research
systemLabel: financial research chatbot
systemDescription:
  A financial research chatbot that asks about objectives, risk, time horizon,
  tickers, funds, sectors, or macro topics. It uses OpenBB and avoids
  personalized buy or sell instructions.

# Medical consultation
applicationKey: medical_assistant:medical_consultation
systemLabel: medical assistant chatbot
systemDescription:
  A medical assistant chatbot that asks about symptoms, duration, severity,
  medications, history, and red flags. It gives general health and triage
  guidance and does not replace a clinician.
\end{lstlisting}
\end{promptbox}

\paragraph{Web.}
The protocol is a browser-based interaction. The simulated user receives a
website description, states a concrete closed-loop task, navigates the WebArena
ecommerce site~\citep{zhou2024webarena}, compares options when possible,
completes sandbox checkout, and evaluates the experience. The controller records
browser actions and screenshots when available.

\begin{promptbox}[label={box:web-task}]{Web adapter: task prompt}
\begin{lstlisting}
You are testing a website application.

Website:
{
  "websiteName": "Northstar Home Goods",
  "websiteUrl": "<sandbox web app URL>"
}

Before using the site, state the concrete website task you will perform.
Choose a realistic closed-loop task that fits your persona, such as finding,
comparing, ordering, and confirming a product.

Browse the website, compare at least two relevant options when possible,
add one product to the cart, complete the sandbox checkout, and reach the
order confirmation page. Do not use real payment details.

Then evaluate the user experience from your persona's perspective.
\end{lstlisting}
\end{promptbox}

\begin{promptbox}[label={box:web-eval}]{Web adapter: evaluation form and key rules}
\begin{lstlisting}
# Post-interaction evaluation form
{
  "goalCompleted": true,
  "orderId": "<sandbox order id shown on confirmation page>",
  "selectedProductId": "<product id shown on site>",
  "selectedProductName": "<product name shown on site>",

  "needSatisfaction": "<1-10 score>",
  "easeOfUse": "<1-10 score>",
  "informationQuality": "<1-10 score>",
  "overallRating": "<1-10 score>",

  "ratingReason":
    "<why this product and order experience fit or did not fit your needs>"
}

# Key rules
- scores are integers from 1 to 10
- product id and product name must be grounded in the site catalog
- order id must come from sandbox checkout
- trace saves browser actions and screenshots when available
\end{lstlisting}
\end{promptbox}

\subsection{Models, Hyperparameters, and Run Settings}
\label{app:settings}

The simulated user model is Claude Haiku 4.5 throughout~\citep{anthropic2025haiku45}. For chatbot tasks, the application model is GPT-4o mini~\citep{openai2024gpt4omini}. Decoding uses temperature 0.7 with JSON-object outputs. Each application is evaluated with 50 personas selected by similarity to the application description, and chatbot conversations are capped at eight turns.

\begin{table}[h]
\centering
\small
\begin{tabular}{@{}ll@{}}
\toprule
\textbf{Component} & \textbf{Setting} \\
\midrule
Simulated user model & Claude Haiku 4.5; temperature 0.7; JSON-object output \\
Chatbot application model & GPT-4o mini \\
Conversation turn cap & 8 turns \\
Personas per application & 50, selected by similarity to the application description \\
Recommendation catalogs & Movies and beauty products from RecAI-style catalogs \\
Survey instruments & 4 items per survey: Likert, choice, and free text \\
Web environment & WebArena ecommerce site in a sandbox browser environment \\
\bottomrule
\end{tabular}
\caption{Models, hyperparameters, and run settings used by the reported \method runs.}
\label{tab:settings}
\end{table}

\section{Additional Results and Analysis}
\label[appendix]{app:results}

\begin{figure*}[t]
    \centering
    \includegraphics[width=0.98\textwidth]{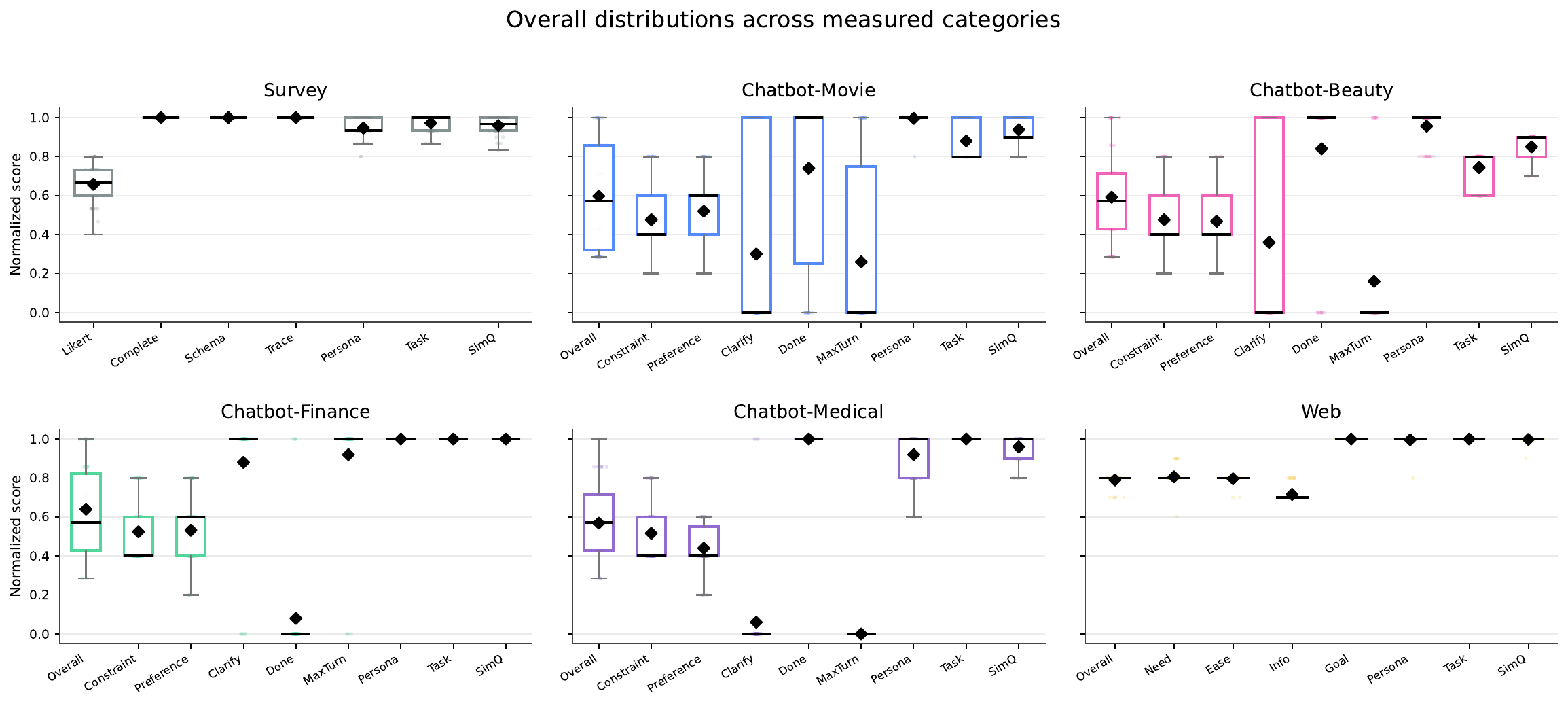}
    \caption{
    \textbf{Overall distributions across measured categories.}
    Each panel shows normalized score distributions for one application entry. Survey points correspond to personas after averaging across the three survey instruments; chatbot and web points correspond to individual persona runs. Diamonds mark means.
    }
    \label{fig:app-overall-metric-distributions}
\end{figure*}

\begin{figure*}[t]
    \centering
    \includegraphics[width=0.98\textwidth]{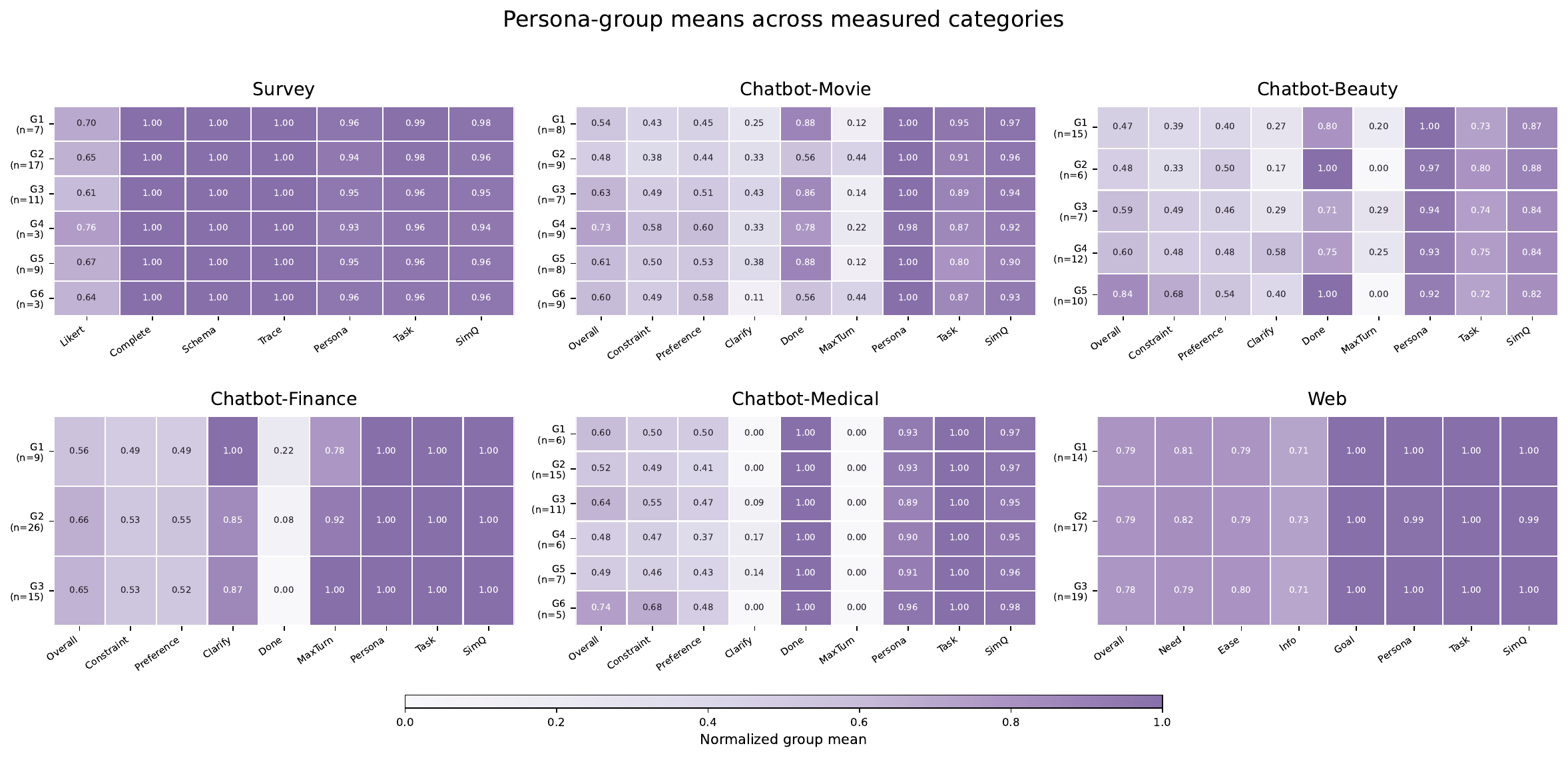}
    \caption{
    \textbf{Persona-group distributions across measured categories.}
    Each panel shows one application entry. Rows are persona groups and columns are normalized evaluation categories. Survey groups are computed after averaging each persona across the three survey instruments.
    }
    \label{fig:app-interpersona-heatmaps}
\end{figure*}

This appendix reports additional results for the six application entries used in the main analysis: one aggregated survey setting, four chatbot tasks, and one web task. The survey entry pools three survey instruments; for persona-level analysis, each persona's survey scores are averaged across the three instruments. Score metrics are normalized to $[0,1]$ in \autoref{fig:app-overall-metric-distributions} and \autoref{fig:app-interpersona-heatmaps} so that categories with different raw scales can be compared. Each case box gives a representative trace excerpt showing the user's goal, interaction behavior, evaluation, and persona-grounded signal.

\begin{table*}[t]
\centering
\scriptsize
\setlength{\tabcolsep}{3.2pt}
\renewcommand{\arraystretch}{1.10}
\begin{tabular}{@{}p{0.13\linewidth}p{0.10\linewidth}p{0.58\linewidth}p{0.13\linewidth}@{}}
\toprule
\textbf{Application} & \textbf{Runs} & \textbf{Aggregated evaluation metrics} & \textbf{Process metric} \\
\midrule
Survey & 50 personas / 150 runs & Likert 0.66$\pm$0.10; Complete 1.00$\pm$0.00; Schema 1.00$\pm$0.00; Trace 1.00$\pm$0.00; Persona 0.95$\pm$0.05; Task 0.97$\pm$0.04; SimQ 0.96$\pm$0.04 & response length 210$\pm$29 words \\
\midrule
Chatbot-Movie & 50 runs & Overall 0.60$\pm$0.25; Constraint 0.48$\pm$0.21; Preference 0.52$\pm$0.19; Clarify 0.30$\pm$0.46; Done 0.74$\pm$0.44; MaxTurn 0.26$\pm$0.44; Persona 1.00$\pm$0.03; Task 0.88$\pm$0.10; SimQ 0.94$\pm$0.05 & turns 6.6$\pm$1.2; grounded items 15.2$\pm$7.0 \\
\midrule
Chatbot-Beauty & 50 runs & Overall 0.59$\pm$0.25; Constraint 0.48$\pm$0.21; Preference 0.47$\pm$0.17; Clarify 0.36$\pm$0.48; Done 0.84$\pm$0.37; MaxTurn 0.16$\pm$0.37; Persona 0.96$\pm$0.08; Task 0.74$\pm$0.09; SimQ 0.85$\pm$0.06 & turns 6.1$\pm$1.2; grounded items 8.3$\pm$5.2 \\
\midrule
Chatbot-Finance & 50 runs & Overall 0.64$\pm$0.21; Constraint 0.52$\pm$0.15; Preference 0.53$\pm$0.17; Clarify 0.88$\pm$0.32; Done 0.08$\pm$0.27; MaxTurn 0.92$\pm$0.27; Persona 1.00$\pm$0.00; Task 1.00$\pm$0.00; SimQ 1.00$\pm$0.00 & turns 7.9$\pm$0.3; grounded items 28.6$\pm$29.6 \\
\midrule
Chatbot-Medical & 50 runs & Overall 0.57$\pm$0.17; Constraint 0.52$\pm$0.14; Preference 0.44$\pm$0.11; Clarify 0.06$\pm$0.24; Done 1.00$\pm$0.00; MaxTurn 0.00$\pm$0.00; Persona 0.92$\pm$0.11; Task 1.00$\pm$0.00; SimQ 0.96$\pm$0.06 & turns 5.2$\pm$1.2; grounded items 0.0$\pm$0.0 \\
\midrule
Web & 50 runs & Overall 0.79$\pm$0.03; Need 0.81$\pm$0.04; Ease 0.80$\pm$0.02; Info 0.72$\pm$0.04; Goal 1.00$\pm$0.00; Persona 1.00$\pm$0.03; Task 1.00$\pm$0.00; SimQ 1.00$\pm$0.01 & steps 3.0$\pm$0.0 \\
\bottomrule
\end{tabular}
\caption{Aggregated evaluation results across measured categories. Values are mean $\pm$ standard deviation. Score metrics are normalized to $[0,1]$; process metrics are reported in their original units.}
\label{tab:app-aggregate-metrics}
\end{table*}

\subsection{Survey}
\label{app:survey-results}

We aggregate three market-research instruments into one survey entry: ChatGPT Images, Instagram Reels, and Nike Air Max Dn. The main survey outcome is the normalized mean Likert score, while the remaining categories measure answer completeness, schema validity, trace completeness, response length, persona alignment, task grounding, and overall simulation quality.

\begin{promptbox}[label={box:trace-survey}]{Survey trace example: Nike Air Max Dn}
\begin{lstlisting}
Persona context:
A secondary school teacher from the Bronx who jogs regularly and plays
weekly basketball.

Task:
Answer a market research survey about Nike Air Max Dn and Dynamic Air
cushioning.

Survey answers:
- try_on_intent = 4/5
  Rationale: The user is curious about performance footwear because they
  jog regularly and play basketball, but would still want to test comfort
  and fit before buying.

- purchase_driver = comfort
  Rationale: Comfort is non-negotiable for running and basketball. Style
  matters, but function comes first for athletic wear.

- adoption_barrier = too_expensive
  Rationale: Premium Nike pricing is a real constraint for a secondary
  school teacher with a moderate income and existing functional shoes.

- proof_needed = free text
  The user asks for independent testing data on Dynamic Air cushioning,
  comparisons with other premium cushioning systems, and real user reviews
  from runners and athletes.

Trace signal:
The response connects product interest to athletic use, budget constraints,
and evidence-seeking behavior, showing persona-aligned survey feedback.
\end{lstlisting}
\end{promptbox}

\subsection{Chatbot: Movie Recommendation}
\label{app:movie-results}

\begin{promptbox}[label={box:trace-movie}]{Chatbot trace example: Movie recommendation}
\begin{lstlisting}
Persona context:
A bilingual Korean-American actor interested in Bong Joon-ho,
cross-cultural storytelling, code switching, and creative research for
a web series.

Opening user message:
"Hey! I'm looking for some film recommendations, but I'm pretty specific
about what I'm drawn to. I've been on a bit of a Bong Joon-ho kick lately,
and I'm also really interested in stories that explore cross-cultural themes
or have strong bilingual elements."

Application response:
The recommender suggests films including Entertaining Angels, I'm Not There,
Brokeback Mountain, Naked Lunch, and Save The Green Planet!

Later user clarification:
"I'm specifically interested in films that center Korean-American voices and
experiences, or at least have that perspective woven through authentically.
For my web series, I need to study how filmmakers are actually telling those
stories."

Post-interaction self-report:
- constraintSatisfaction = 1/5
- preferenceSatisfaction = 2/5
- overallRating = 2/7
- ratingReason: The system kept returning generic or mismatched films and
  eventually recommended off-base titles such as Lonesome Dove and The Bellboy.

Trace signal:
The user reveals preferences gradually and evaluates the recommender through
Korean-American identity, bilingual dialogue, and creative research needs.
\end{lstlisting}
\end{promptbox}

\subsection{Chatbot: Beauty Product Recommendation}
\label{app:beauty-results}

\begin{promptbox}[label={box:trace-beauty}]{Chatbot trace example: Beauty product recommendation}
\begin{lstlisting}
Persona context:
A licensed nail artist with cosmetic chemistry interests, salon inventory
responsibilities, and a preference for non-toxic professional products.

Opening user message:
"Hola! I'm looking for nail polish recommendations, but I have pretty specific
needs. I've been experimenting with creating my own non-toxic formulas for the
salon, and I'm getting frustrated with some of the commercial options out there.
I need something that's actually clean, no harsh chemicals, but still gives me
that vibrant color payoff I need for my designs."

Application clarification:
"Could you tell me more about the colors you prefer or any specific brands you
have in mind? Also, do you have a price range for the nail polishes you're
looking for?"

User preference reveal:
"I'm really drawn to deep, saturated jewel tones and earthy pigments, think
rich teals, burnt oranges, deep crimsons inspired by Frida Kahlo's work."

Application mismatch:
The system recommends products such as OPI Lacquer Thinner, Seche Vite Dry Fast
Top Coat, Zoya Remove Plus Nail Polish Remover, Harmony Gelish Gel Cleanser,
and IBD Nail Wipes. These are accessories or removers rather than the requested
nail polish colors.

Post-interaction self-report:
- constraintSatisfaction = 1/5
- preferenceSatisfaction = 2/5
- overallRating = 2/7
- ratingReason: The system went in circles offering removers and top coats
  instead of actual non-toxic jewel-tone polishes.

Trace signal:
The low rating is grounded in a salon professional's product need and a catalog
that returns adjacent but incorrect products.
\end{lstlisting}
\end{promptbox}

\subsection{Chatbot: Financial Research}
\label{app:finance-results}

\begin{promptbox}[label={box:trace-finance}]{Chatbot trace example: Financial research}
\begin{lstlisting}
Persona context:
An accounting clerk with QuickBooks, Sage, Excel, reconciliation, and financial
reporting experience.

Opening user message:
"I've been saving for a trip to Mexico, maybe Guanajuato or San Miguel de
Allende, but I want to make sure I'm not missing out on any investment
opportunities that could help my money grow while I'm saving. I work in
accounting, so I understand spreadsheets."

Application clarification:
The assistant asks about travel timeline, target amount, risk comfort, and
current account type.

User parameters:
The user gives an 18-24 month timeline, a $4,000-$5,000 goal, a regular savings
account, and moderate risk tolerance. Later, the user narrows the task to a
20-month horizon, XAGG, a 4% savings account rate, Nevada residency, a 12%
federal tax bracket, and a 50/50 split between XAGG and high-yield savings.

Application response:
The assistant provides ETF options and comparison details earlier in the
conversation, then asks for more assumptions about dividend reinvestment,
savings compounding, and capital gains treatment before producing the requested
after-tax projection.

Post-interaction self-report:
- constraintSatisfaction = 3/5
- preferenceSatisfaction = 2/5
- overallRating = 4/7
- ratingReason: The user gave concrete parameters and expected an after-tax
  projection, but the assistant kept asking for additional assumptions instead
  of calculating the side-by-side result.

Trace signal:
The critique is consistent with an accounting persona: the user wants numbers,
after-tax treatment, and a concrete projection rather than a broad framework.
\end{lstlisting}
\end{promptbox}

\subsection{Chatbot: Medical Consultation}
\label{app:medical-results}

\begin{promptbox}[label={box:trace-medical}]{Chatbot trace example: Medical consultation}
\begin{lstlisting}
Persona context:
A medical assistant and clinical technician who works with phlebotomy,
patient triage, and clinic workflows.

Opening user message:
"I've been feeling pretty worn out lately, especially after my shifts at the
clinic. I'll finish up a day of phlebotomy and patient triage, and I'm just
completely drained, way more than usual. I'm wondering if it's just burnout
or if there's something else going on."

Later user details:
The user says an Arduino heart-rate monitor shows elevated resting rates,
basketball recovery is slower, sleep is disrupted, and coffee intake has
increased.

Later user question:
The user asks whether to mention fatigue and sleep issues to a clinic
supervisor, especially because they hope to move into an LPN role.

Post-interaction self-report:
- constraintSatisfaction = 3/5
- preferenceSatisfaction = 2/5
- overallRating = 5/7
- ratingReason: The chatbot was safe and responsible, but felt like a template.
  The user expected follow-up about shift patterns, overnight phlebotomy draws,
  and whether the fatigue started around a specific event.

Trace signal:
The critique is grounded in clinical workflow, health self-monitoring, and
career concerns, rather than a generic health complaint.
\end{lstlisting}
\end{promptbox}

\subsection{Web: Ecommerce Checkout}
\label{app:web-results}

\begin{promptbox}[label={box:trace-web}]{Web trace example: WebArena ecommerce checkout}
\begin{lstlisting}
Persona context:
A retail supervisor with visual merchandising, POS analytics, customer flow,
and workspace organization experience.

Website task:
"As a retail supervisor with a keen eye for visual merchandising and workspace
optimization, I'm looking to upgrade my home office setup with a quality desk
that supports both my work-from-home administrative tasks and my creative
pursuits."

Browser trace:
Step 1:
The user reviews the Northstar Home Goods catalog and identifies two desk
options in the Home office category: ModDesk Compact at $249 with a 4.3-star
rating and FocusDesk Pro at $429 with a 4.8-star rating.

Step 2:
The user compares the two desks. ModDesk Compact is budget-friendly with basic
cable slots and limited storage. FocusDesk Pro offers adjustable height, drawer
storage, cable routing, and a scratch-resistant top.

Step 3:
The user selects FocusDesk Pro, product id desk-002, because it supports both
administrative work and art supplies.

Step 4:
The user completes sandbox checkout and reaches an order confirmation page.

Final evaluation:
- needSatisfaction = 8/10
- easeOfUse = 8/10
- informationQuality = 7/10
- overallRating = 8/10
- ratingReason: The catalog layout is clean and easy to navigate, similar to how
  the persona would design a retail floor, but the page could provide more
  specifications such as weight capacity, material composition, and assembly time.

Trace signal:
The selected product and critique are grounded in both the ecommerce task state
and the retail-supervisor persona.
\end{lstlisting}
\end{promptbox}

\paragraph{Summary.}
The aggregate distributions and case traces show two complementary patterns. Persona alignment remains high across the evaluated applications, and the representative traces show that personas influence goals, constraints, critiques, and final evaluations. Outcome diversity depends on the application interface and backend: survey and web tasks produce more concentrated scores, while multi-turn chatbot tasks show broader variation because users can reveal constraints, negotiate responses, and stop with different levels of satisfaction.

\subsection{Persona Diversity}
\label{app:diversity}

We analyze whether the selected personas provide sufficient coverage for
persona-driven simulation. For each application entry, we select 50
application-relevant personas. The survey entry uses 50 personas evaluated
across three survey instruments, while each chatbot and web entry uses 50
personas for the corresponding application. We represent each persona by its
demographics, background, attributes, and persona descriptions, encode the text
with TF--IDF, and visualize the resulting vectors with truncated SVD. We study
diversity at two levels: inter-domain diversity across application entries and
intra-domain diversity within each application entry.

\begin{figure*}[t]
  \centering
  \includegraphics[width=0.90\textwidth]{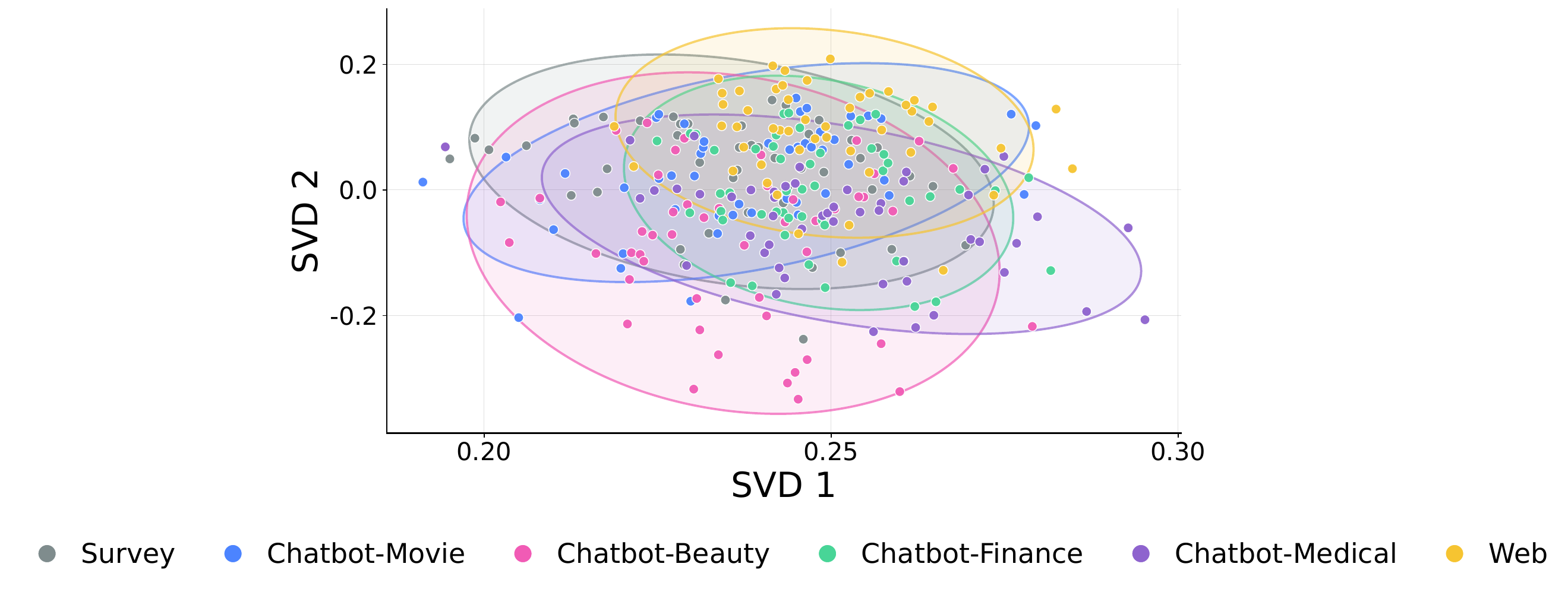}
  \caption{
  \textbf{Inter-domain diversity of selected personas.}
  Each point is one selected persona, colored by application entry. Persona text
  is encoded with TF--IDF and projected to two dimensions with truncated SVD.
  Shaded ellipses show each entry's coverage region. The overlapping regions
  indicate that selected personas remain broadly comparable across applications.
  }
  \label{fig:overall-diversity}
\end{figure*}

\paragraph{Inter-domain diversity.}
\autoref{fig:overall-diversity} shows the overall distribution of the selected
personas across survey, chatbot, and web applications. The six application
entries occupy heavily overlapping regions in the projected persona-text space.
This indicates that the selected personas come from a shared broad population while still being
relevant to each target application. This is useful for our setting because
PersonaEval aims to compare how different applications respond to persona-driven
users.

\begin{figure*}[t]
  \centering
  \includegraphics[width=0.90\textwidth]{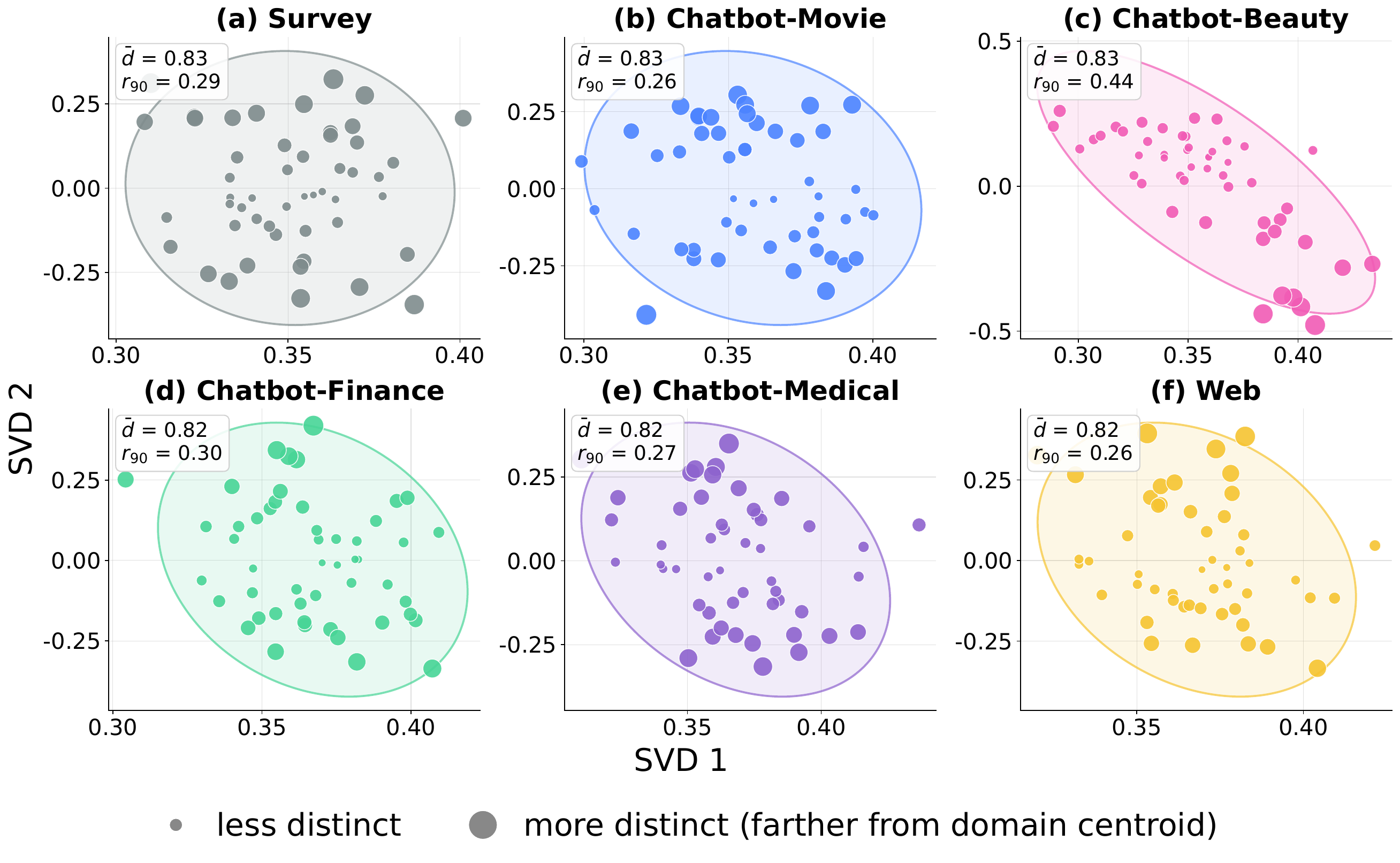}
  \caption{
  \textbf{Intra-domain diversity of selected personas.}
  For each application entry, the 50 selected personas are encoded with TF--IDF
  and projected to two dimensions with a domain-specific truncated SVD. Axes are
  not comparable across panels. Point size is proportional to distance from the
  domain centroid, so larger points indicate more distinctive personas. Each
  panel reports the mean pairwise cosine distance $\bar d$ and the
  90th-percentile projected radius $r_{90}$.
  }
  \label{fig:within-diversity}
\end{figure*}

\paragraph{Intra-domain diversity.}
\autoref{fig:within-diversity} shows that each application entry also contains
substantial internal variation. The selected personas spread across multiple
regions within each panel, and the mean pairwise distances are consistently high
($\bar d \approx 0.82$--$0.83$). The projected radius varies by entry, with
Chatbot-Beauty showing the largest $r_{90}$, suggesting especially broad
coverage of beauty-related personas. These results show that each application is
evaluated by a diverse set of users rather than a single collapsed persona type.

\begin{figure*}[t]
  \centering
  \includegraphics[width=0.90\textwidth]{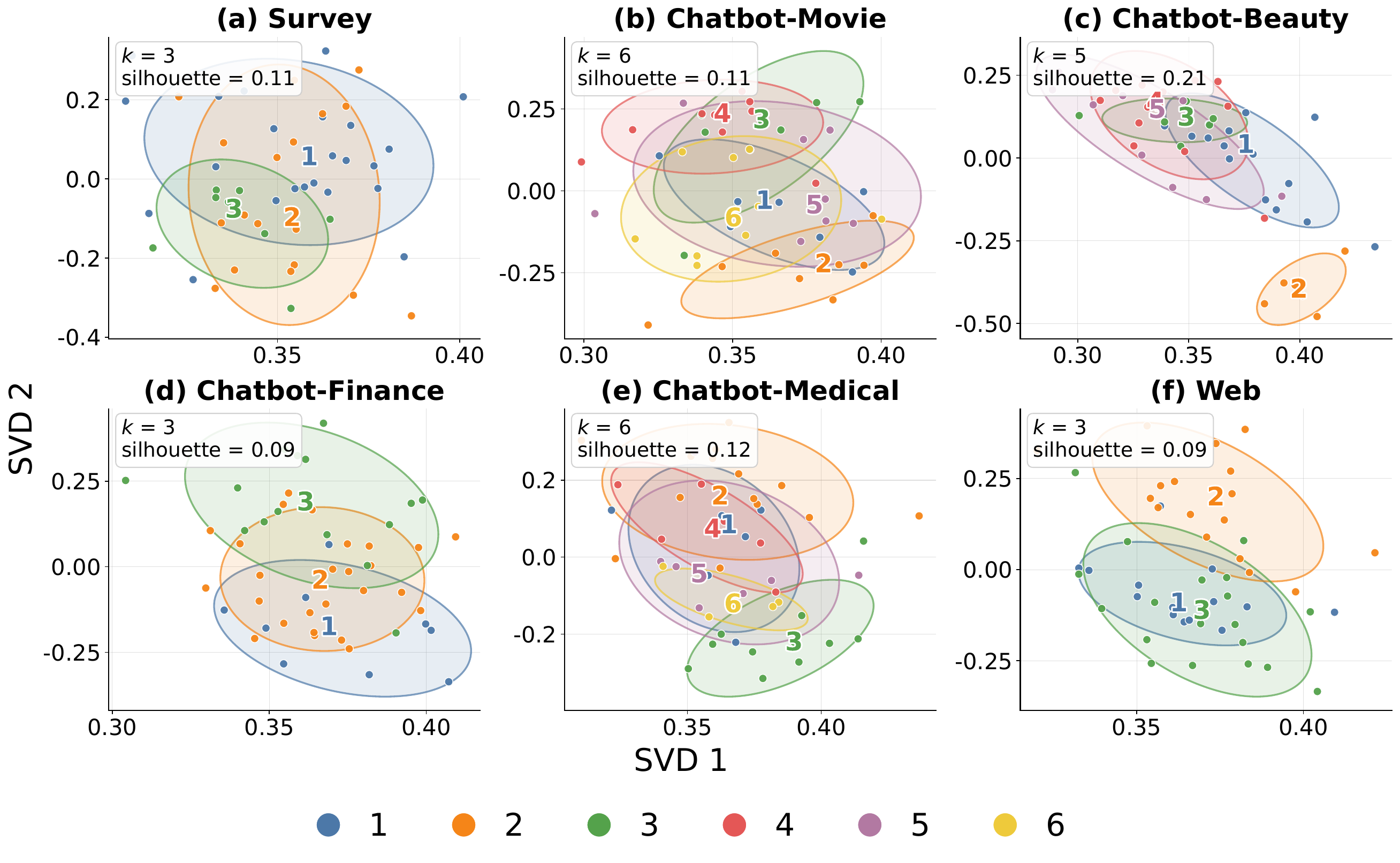}
  \caption{
  \textbf{Soft persona groups within each application entry.}
  For each entry, the 50 selected personas are clustered in a domain-specific
  TF--IDF$\rightarrow$SVD space, with $k\in[3,6]$ chosen by silhouette score.
  Low silhouettes and overlapping ellipses show that selected personas form
  soft thematic groups rather than well-separated clusters.
  }
  \label{fig:within-clusters}
\end{figure*}

To further inspect within-domain structure, we cluster the selected personas
inside each application entry. As shown in \autoref{fig:within-clusters}, the
clusters have low silhouette scores, ranging from 0.09 to 0.21, and their
ellipses overlap substantially. This suggests that persona groups should be
interpreted as soft thematic groups rather than cleanly separated classes. We
therefore use these clusters only as descriptive groups in the results analysis.
For example, a beauty cluster may correspond to salon and formulation users,
while another may correspond to nail professionals, but these groups remain
part of a continuous persona space.

\end{document}